# First-principles study of the thermoelectric properties of quaternary tetradymite BiSbSeTe$_2$

Z. Z. Zhou, H. J. Liu[*], D. D. Fan, B. Y. Zhao, C. Y. Sheng, G. H. Cao, S. Huang

*Key Laboratory of Artificial Micro- and Nano-Structures of Ministry of Education and School of Physics and Technology, Wuhan University, Wuhan 430072, China*

The electronic and phonon transport properties of quaternary tetradymite BiSbSeTe$_2$ are investigated using first-principles approach and Boltzmann transport theory. Unlike the binary counterpart Bi$_2$Te$_3$, we obtain a pair of Rashba splitting bands induced by the absence of inversion center. Such unique characteristic could lead to a large Seebeck coefficient even at relatively higher carrier concentration. Besides, we find an ultralow lattice thermal conductivity of BiSbSeTe$_2$, especially along the interlayer direction, which can be traced to the extremely small phonon relaxation time mainly induced by the mixed covalent bonds. As a consequence, a considerably large *ZT* value of ~2.0 can be obtained at 500 K, indicating that the unique lattice structure of BiSbSeTe$_2$ caused by isoelectronic substitution could be an advantage to achieving high thermoelectric performance.

## 1. Introduction

With the increasing global energy crisis and environmental pollution, it is necessary to develop efficient thermoelectric devices which can convert waste heat into electricity and vice versa. The performance of a thermoelectric material is determined by the dimensionless figure of merit $ZT = S^2\sigma T/(\kappa_e + \kappa_l)$, where $S$ is the Seebeck coefficient, $\sigma$ is the electrical conductivity, $T$ is the absolute temperature, $\kappa_e$ and $\kappa_l$ are the electronic thermal conductivity and the lattice thermal conductivity, respectively. However, it is usually difficult to achieve a high *ZT* value because almost all of the thermoelectric transport coefficients are more or less coupled with each other. There are some effective ways to improve the *ZT* value, such as employing low-dimensional materials [1], forming nanostructures [2], and using superlattice structures [3]. Of course, materials used as thermoelectric devices should have suitable band gap and low lattice thermal conductivity as a premise.

---

[*] Author to whom correspondence should be addressed. Electronic mail: phlhj@whu.edu.cn

Among many bulk thermoelectric materials, the tetradymite compounds have been extensively studied for a long time. In particular, the binary tetradymite $Bi_2Te_3$ with a *ZT* value of ~1.0 [4] are known as one of the best thermoelectric materials operated in the vicinity of room temperature. However, such a *ZT* is still far from the target value of 3.0 for commercial applications. To further enhance its thermoelectric performance, one effective approach is the band engineering via isoelectronic substitution [5]. Indeed, a lot of experimental works have been focused on the thermoelectric properties of the ternary and quaternary tetradymite compounds. For example, Poudel *et al.* [6−8] prepared the *p*-type bulk Bi-Sb-Te alloys via high energy ball milling and hot pressed technique, and found a *ZT* of ~1.4 at 373 K. Wang *et al.* [9] reported that the *n*-type $(Bi_{0.95}Sb_{0.05})_2(Te_{0.85}Se_{0.15})_3$ shows a *ZT* value of 1.0 at 460 K. Zhang *et al.* [10] synthesized the BiSbTeSe alloys with *x*-wt% ZnSb (*x* = 0, 0.3, 0.5 and 1.0) addition, and a maximum *ZT* value of 1.13 could be obtained at 410 K for the sample with *x* = 0.5. On the theoretical side, density functional calculations were performed by Ryu *et al.* [11] to study the band structures of Sb/Se-doped ternary compounds $(Bi_{1−x}Sb_x)_2Te_3$ and $Bi_2(Te_{1−y}Se_y)_3$, where the band gaps were found to be increased by Sb/Se doping. Using first-principles calculations, Shi *et al.* [12] suggested that the complex non-parabolic band structures of $Bi_2Te_2Se$ could be favorable to achieve high thermoelectric performance. Besides, Devender *et al.* [13] demonstrated that sulfur doping could lead to simultaneous multifold increases in *S* and *σ* for the *n*-type $Bi_2Te_2Se$ due to profound changes in the band structure. In contrast to the extensive studies on the ternary compounds, theoretical investigation on the thermoelectric properties of quaternary tetradymites is less known and a complete understanding is quite necessary. Unlike the binary counterpart $Bi_2Te_3$, the ternary and quaternary tetradymite compounds may not have the inversion center so that the Rashba effect [14] could be induced by the spin-orbital coupling (SOC). Moreover, Wu *et al.* [15] suggested that the Rashba effect in the inversion asymmetric BiTeI could be beneficial to enhance the thermoelectric performance. It is thus natural to ask whether it is also the case for the quaternary tetradymite compounds.

In this work, the electronic, phonon, and thermoelectric transport properties of quaternary tetradymite $BiSbSeTe_2$ have been studied by using first-principles calculations and Boltzmann transport theory. We find that the Rashba effect and polarization field [16] induced by the absence of inversion center have obvious influences on the electronic band structure and phonon dispersion relations, respectively. Compared with the binary and ternary tetradymites, considerably larger *ZT* values can be obtained in such quaternary compound.

## 2. Computational details

The electronic properties of $BiSbSeTe_2$ have been investigated by using the projector-augmented wave (PAW) method [17] within the framework of density functional theory (DFT) [18, 19], as implemented in the Vienna *ab-initio* simulation package (VASP) [20]. The exchange-correlation functional is Perdew-Burke-Ernzerhof (PBE) with generalized gradient approximation (GGA) [21]. To accurately predict the band gap, the hybrid density functional in the form of Heyd-Scuseria-Ernzerhof (HSE06) is also considered [22−24]. The cutoff energy for the plane wave basis is set to 500 eV and the energy convergence threshold is $10^{-6}$ eV. The Brillouin zone is sample with a $13\times13\times13$ Monkhorst-Pack *k* mesh grids. The van der Waals (vdW) interaction [25] and the SOC are explicitly included in the calculations.

The electronic transport coefficients are evaluated by using the Boltzmann transport theory with the relaxation time approximation [26]. In this approach, the Seebeck coefficient, the electrical conductivity, and the electronic thermal conductivity are respectively given by

$$S = \frac{ek_B}{\sigma} \int d\varepsilon \left(-\frac{\partial f_0}{\partial \varepsilon}\right) \Xi(\varepsilon) \frac{\varepsilon - \mu}{k_B T}, \quad (1)$$

$$\sigma = e^2 \int d\varepsilon \left(-\frac{\partial f_0}{\partial \varepsilon}\right) \Xi(\varepsilon), \quad (2)$$

$$\kappa_e = \frac{1}{T} \int d\varepsilon \left(-\frac{\partial f_0}{\partial \varepsilon}\right) \Xi(\varepsilon)(\varepsilon - \mu)^2 - TS^2\sigma, \quad (3)$$

where $k_B$ is the Boltzmann constant, $\mu$ is the chemical potential, and $f_0 = \frac{1}{e^{(\varepsilon-\mu)/k_B T} + 1}$ is the Fermi-Dirac function. The so-called transport distribution function $\Xi(\varepsilon) = \sum_{\vec{k}} \vec{v}_{\vec{k}} \vec{v}_{\vec{k}} \tau_{\vec{k}}$ can be obtained from the band structure, where $\tau_{\vec{k}}$ and $\vec{v}_{\vec{k}}$ are the relaxation time and group velocity, respectively.

The phonon dispersion relations of BiSbSeTe$_2$ are obtained by using the finite displacement method as implemented in the Phonopy package [27]. A $4\times4\times4$ and $3\times3\times3$ supercells are employed for the calculation of the second- and third-order interatomic force constants (IFCs), respectively. In addition, the interactions up to the ninth nearest neighbors are considered to obtain convergent results for the anharmonic IFCs. The phonon transport properties can be estimated by solving the Boltzmann transport equation as implemented in the so-called ShengBTE code [28]. A $27\times27\times27$ $q$-mesh is adopted to ensure that the lattice thermal conductivity is converged.

## 3. Results and discussion

As can be seen from Figure 1(a), bulk BiSbSeTe$_2$ crystallizes in a rhombohedral structure with space group $R\bar{3}m$. Similar to Bi$_2$Te$_3$, bulk BiSbSeTe$_2$ constitutes stacks of quintuple layers (QLs) by covalent bonds in the sequence of Te1-Bi-Se-Sb-Te2, and the QLs are bonded together with weak vdW interactions. The lattice parameters of the primitive cell are $a_0$ = 10.34 Å and $\alpha$ = 23.95º, which correspond to $a$ = 4.29 Å and $c$ = 30.12 Å in a hexagonal unit cell as shown in Fig. 1(b). Compared with the binary $Bi_2Te_3$ [29], there is no inversion center in the quaternary $BiSbSeTe_2$ caused by the inequivalent positions of Bi and Sb atoms. Besides, there exist four different bond lengths in the quaternary $BiSbSeTe_2$ with Te−Bi (2.996 Å), Bi−Se (3.149 Å), Se−Sb (3.156 Å), and Sb−Te (2.979 Å), and they are quite different from the Bi−Te bonds (3.082 Å and 3.262 Å) in the binary $Bi_2Te_3$. It is expected that such unique structure characteristics may have significant effects on the electronic and phonon transport properties of $BiSbSeTe_2$.

Figure 2(a) plots the PBE-calculated band structure of $BiSbSeTe_2$, where the labelled conduction band minimum (CBM) and valence band maximum (VBM) give a band gap of 0.20 eV. However, we should note that the CBM and VBM of several tetradymite compounds actually do not locate at those high-symmetry lines [25, 30]. The real band gap is calculated to be 0.15 eV by a careful search in the whole Brillouin zone using very dense $k$ mesh. If we focus on the conduction bands, we observe a pair of Rashba splitting bands around the $\Gamma$ point, which is mainly caused by the absence of inversion center in the lattice structure of $BiSbSeTe_2$. It was found that at the same carrier concentration, the Fermi level is relatively lower for materials with the Rashba effect, which could lead to a higher Seebeck coefficient [15]. Fig. 2(b) plots the HSE-calculated band structure of $BiSbSeTe_2$. Compared with the PBE result, we see that the valence valley along the $ZF$ direction is obviously downshifted so that the band gap is increased to 0.27 eV. Moreover, there are three valence valleys with almost identical energy along the $ZF$, $F\Gamma$ and $\Gamma L$ directions, and such kind of multi-valley structures usually lead to large Seebeck coefficient [31]. To have a better understanding of the Rashba effect, the enlarged view of the splitting conduction and valence bands are shown in Fig. 2(c) and 2(d), respectively. The Rashba parameter $\alpha_R$ can be defined as $\alpha_R = 2E_0/k_0$ by taking the parabolic approximation [32], where the Rashba energy $E_0$ and Rashba momentum $k_0$ are fitted from first-principles calculations. The calculated $\alpha_R$ are 2.7 and 1.7 eV·Å for the splitting conduction and valence bands, respectively. Such large values indicate stronger Rashba effect, which could have important influence on the electronic transport properties of BiSbSeTe$_2$.

We now discuss the electronic transport coefficients of BiSbSeTe$_2$. Within the framework of Boltzmann transport theory, both the electrical conductivity $\sigma$ and the electronic thermal conductivity $\kappa_e$ depend on the relaxation time $\tau$. Here we use the deformation potential (DP) theory [33] to evaluate $\tau$, which can be expressed as

$$\tau = \frac{2\sqrt{2\pi}C\hbar^4}{3\left(k_B T m^*_{dos}\right)^{3/2} E^2}. \qquad (4)$$

In this formula, $m^*_{dos}$ is the density of states (DOS) effective mass, $C$ is the elastic constant, and $E$ is the deformation potential constant which represents the shift of band edges (VBM and CBM) per unit strain. The calculated room temperature relaxation time and the corresponding parameters are summarized in Table 1, where the results for the in-plane and out-of-plane are both shown. It is obvious that the relaxation time of *p*-type carriers is larger than that of *n*-type carriers, which is mainly caused by the difference in their DOS effective mass and suggests that *p*-type BiSbSeTe$_2$ may have better thermoelectric performance. We thus focus on the *p*-type system in the following discussions. Moreover, we find that the relaxation time of BiSbSeTe$_2$ exhibits strong anisotropy, which is larger for the *x*-direction (in-plane) than that for the *z*-direction (out-of-plane). This is reasonable since the interlayer vdW interactions and mixed covalent bonds could greatly enhance the carrier scattering.

In Figure 3, we show the room temperature Seebeck coefficient ($S$), the electrical conductivity ($\sigma$), the electronic thermal conductivity ($\kappa_e$), and the power factor ($PF = S^2\sigma$) of BiSbSeTe$_2$, plotted as a function of hole concentration along the *x*-and *z*-directions. Fig. 3(a) indicates that the Seebeck coefficients in the *z*-direction are slightly higher than that in the *x*-direction, and they are almost identical to each other with increasing carrier concentration. At the carrier concentration of $10^{19}$ cm$^{-3}$, the absolute values of Seebeck coefficient can reach ~280 μV/K, which is even higher than those of many good thermoelectric material such as SnSe (~200 µV/K) [34]. Such high Seebeck coefficient is believed to be induced by the multi-valley band structures as indicated in Fig. 2. On the other hand, the electrical conductivity (Fig. 3(b)) and the electronic thermal conductivity (Fig. 3(c)) exhibit strong direction dependence, which is consistent with the anisotropic carrier relaxation time discussed above. With the increase of carrier concentration, the electrical conductivity and the Seebeck coefficient show a reversed behavior, indicating that a compromise must be taken to maximize the power factor. Due to the Rashba splitting band structure, one can find relatively low chemical potential at the optimized carrier concentration of $1.6 \times 10^{19} \text{cm}^{-3}$. Such effect could lead to large Seebeck coefficient and electrical conductivity simultaneously [15]. As a result, the power factor exhibited in Fig. 3(d) can be optimized to $1.49 \times 10^{-2} \text{ W}/\text{mK}^2$ for $p$-type system in the $x$-direction, which is almost five times as much as that of $Bi_2Te_3$ [25].

To obtain the lattice thermal conductivity, we first calculate the phonon dispersion relations of $BiSbSeTe_2$. In ionic crystals and polar semiconductors, the optical branches always display the splitting of TO and LO around the $\Gamma$ point due to the polarization field [35]. Considering that $BiSbSeTe_2$ exhibits asymmetric lattice structures, we compare the phonon spectrums in Figure 4(a) and 4(b) with and without the polarization field, respectively. The polarization field is described by the dielectric constants and Born effective charges, which can be calculated by using density functional perturbation theory (DFPT) [36]. It can be clearly seen in Fig. 4(a) that the optical branches display a large splitting of about 39.5 cm$^{-1}$ in the intermediate frequency region around the $\Gamma$ point, where the branches are degenerate without the polarization field (as circled in Fig. (b)). In addition, the maximum phonon frequency is smaller than 180 cm$^{-1}$, which is comparable to those of good thermoelectric materials such as $Bi_2Te_3$ (150 cm$^{-1}$) [37] and SnSe (186 cm$^{-1}$) [38], suggesting the lower thermal conductivity of $BiSbSeTe_2$. Fig. 4(c) shows the phonon DOS (PDOS) of $BiSbSeTe_2$ as a function of frequency, where we see that the acoustic phonon modes are dominated by the Bi and Te atoms. In the frequency region of 0~60 cm$^{-1}$, we find that the PDOS of Bi (Sb) obviously hybridize with that of Te atoms while it is much less for that of Bi (Sb) and Se atoms, indicating a large vibrational mismatch between them. Such observation suggests that the acoustic and low frequency optical phonons may be easily scattered, which could greatly limit their contribution to the heat transport [39, 40]. Fig. 4(d) shows the temperature dependence of the lattice thermal conductivity of BiSbSeTe$_2$. It is obvious that the out-of-plane one ($\kappa_z$) is much smaller than the in-plane one ($\kappa_x$), which can be attributed to the mixed covalent bonds and vdW interaction between adjacent QLs. Note that the $\kappa_x$ of BiSbSeTe$_2$ and Bi$_2$Te$_3$ are almost the same at 300 K, while the $\kappa_z$ in BiSbSeTe$_2$ (0.35 W/mK) is much smaller than that of Bi$_2$Te$_3$ (1.0 W/mK) [41]. Such small values of thermal conductivity suggest that BiSbSeTe$_2$ could have favorable thermoelectric performance. Considering the fact that the system exhibit higher *PF* in the *x*-direction while lower $\kappa_l$ in the *z*-direction, it is reasonable to expect that the in-plane *ZT* value is comparable to the out-of-plane one in the quaternary BiSbSeTe$_2$, which is quite different from that of many binary tetradymite compounds [42]. Figure 5 displays the accumulative lattice thermal conductivity of BiSbSeTe$_2$ as a function of phonon mean free path (MFP) at three typical temperatures. We see that $\kappa_l$ can be obviously reduced in the *x*-direction (*z*-direction) at 300 K if the sample size of BiSbSeTe$_2$ is smaller than 55 nm (28 nm). As temperature is increased, the cutoff MFPs become decreasing, which are 30 nm (18 nm) and 16 nm (11 nm) for 500 K and 800 K, respectively. The intrinsically short MFPs of BiSbSeTe$_2$ limit the potential for reducing the lattice thermal conductivity by conventional nanostructuring.

To understand the intrinsically low lattice thermal conductivity found in BiSbSeTe$_2$ (especially for the out-of-plane one), we plot in Figure 6(a) and 6(b) the room temperature phonon group velocity and relaxation time, respectively. It can be seen that relatively lower phonon group velocity can be found in the intermediate-frequency region around 80 cm$^{-1}$, which is consistent with the flat phonon branches shown in Fig. 4(a). It is noted that the phonon group velocity of BiSbSeTe$_2$ and Bi$_2$Te$_3$ are similar to each other for the LA and TA branches [43]. The lower lattice thermal conductivity of BiSbSeTe$_2$ should be thus caused by its relatively smaller phonon relaxation time. Compared with that of Bi$_2$Te$_3$ (Fig. 6(c)), we see that the acoustic phonon relaxation time of BiSbSeTe$_2$ is one order of magnitude smaller. Considering the different atomic configuration of Bi$_2$Te$_3$ and BiSbSeTe$_2$, it is expected that the mass difference or mixed covalent bonds could play an important role in determining the phonon transport properties. To go into the details, we replot in Fig. 6(d) the phonon relaxation time of BiSbSeTe$_2$ where the mass difference is removed. In another word, we artificially set Bi (Te) and Sb (Se) within the BiSbSeTe$_2$ to have the same atomic mass but with distinct chemical identity. The resulting phonon relaxation time is almost the same as that of Fig. 6(b), which indicates that the mass difference has little effect in the phonon transport. On the other hand, the phonon relaxation time shown in Fig. 6(d) decreases a lot when compared with Fig. 6(c), while the only difference between them is the mixed covalent bonds with different strength. Similar results can be found for other temperatures. It is thus safe to conclude that the ultralow lattice thermal conductivity of BiSbSeTe$_2$ in the interlayer direction is induced by the mixed covalent bonds which greatly increase the phonon scattering rate.

With all the transport coefficients available, we can now evaluate the thermoelectric performance of BiSbSeTe$_2$. Figure 7(a) shows the calculated *ZT* value as a function of hole concentration at room temperature. It can be seen that both the *x*- and *z*-direction exhibit good thermoelectric performance, and the highest *ZT* of 1.24 can be achieved along the *z*-direction at optimized carrier concentration of $1.6 \times 10^{19}$ cm$^{-3}$. Such a high *ZT* value could be attributed to the extremely low thermal conductivity in the interlayer direction. The temperature dependence of the *ZT* value for *p*-type system is also shown in Fig. 7(b), where we find the highest *ZT* value of ~2.0 can obtained at 500 K. The optimized hole concentration is $3.3 \times 10^{19} \text{cm}^{-3}$, as indicated in Table 2 with the corresponding transport coefficients. Such a *ZT* value exceeds the best of those of the binary [44] and ternary [9, 45] tetradymite compounds, suggesting the very promising material design principle via isoelectronic substitution to form unique lattice structure.

## 4. Summary

Using first-principles calculations and Boltzmann theory for both electrons and phonons, we present a comprehensive study on the thermoelectric properties of quaternary tetradymite BiSbSeTe$_2$. It is found that a band gap of 0.27 eV can be obtained by choosing appropriate hybrid functional for the exchange-correlation energy. Due to the absence of inversion center, there are obvious Rashba splitting bands around the Fermi level, which can lead to a large power factor. On the other hand, we find that the polarization field could cause a large splitting of the optical branches around the $\Gamma$ point in the phonon spectrum. Detailed analyses of the phonon group velocity and relaxation time suggest that the extremely low lattice thermal conductivity along the *z*-direction is induced by the mixed covalent bonds. At an optimized hole concentration of $3.3 \times 10^{19} \text{cm}^{-1}$, the BiSbSeTe$_2$ compound exhibits superior thermoelectric performance with a maximum *ZT* value of ~2.0 at 500 K. Our theoretical work demonstrates that good thermoelectric properties can be achieved in the inversion asymmetric tetradymite compounds with strong Rashba effect and mixture of various chemical bonds.

## Acknowledgements

We thank financial support from the National Natural Science Foundation (Grant Nos. 11574236 and 51772220). The numerical calculations in this work have been done on the platform in the Supercomputing Center of Wuhan University.

**Table 1** Elastic modulus $C$, DOS effective mass $m^*$, deformation potential constant $E$, and relaxation time $\tau$ of BiSbSeTe$_2$ compound at room temperature.

| Direction | Carriers | $C$ (eV/ Å$^3$) | $m^*$ (m$_e$) | $E$ (eV) | $\tau$ (fs) |
|---|---|---|---|---|---|
| $x$ | Electron | 0.67 | 0.42 | −15.87 | 53 |
|   | Hole | 0.67 | 0.35 | −14.88 | 82 |
| $z$ | Electron | 0.79 | 0.42 | −31.32 | 16 |
|   | Hole | 0.79 | 0.35 | −29.96 | 24 |

**Table 2** The optimized $ZT$ values of $p$-type BiSbSeTe$_2$ along the $x$- and $z$-directions at 500 K. The corresponding carrier concentration and transport coefficients are also listed.

|   | $n$ (10$^{19}$ cm$^{-3}$) | $S$ (μV/K) | $\sigma$ (S/cm) | $S^2\sigma$ (10$^{-3}$ W/mK$^2$) | $\kappa_e$ (W/mK) | $\kappa_l$ (W/mK) | $ZT$ |
|---|---|---|---|---|---|---|---|
| $x$ | 4.6 | 230 | 1912 | 10.1 | 2.23 | 1.34 | 1.6 |
| $z$ | 3.3 | 267 | 1190 | 8.5 | 1.44 | 0.21 | 2.0 |

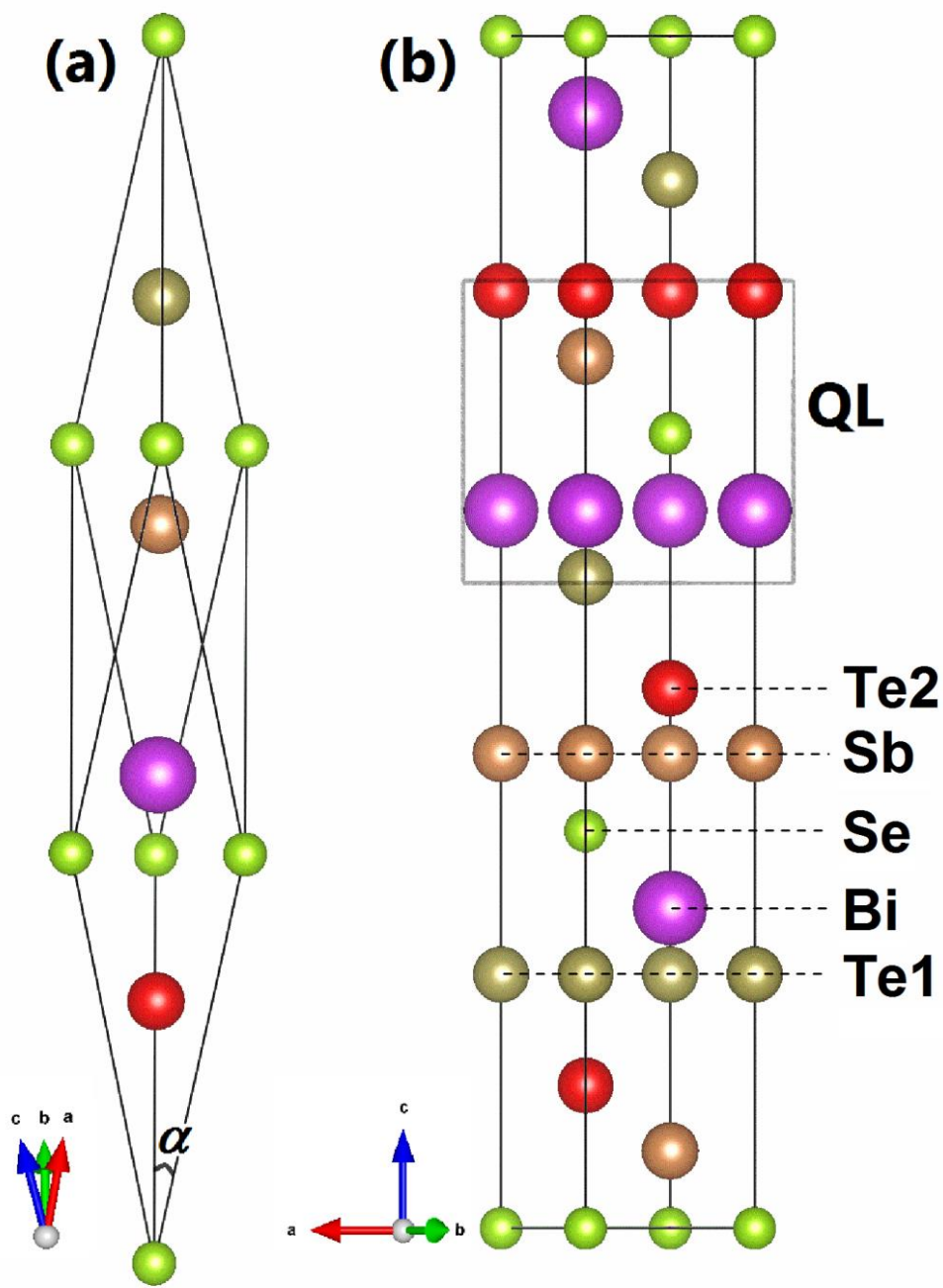

**Figure 1.** The crystal structure of BiSbSeTe$_2$ shows (a) rhombohedral unit cell, and (b) hexagonal conventional cell.

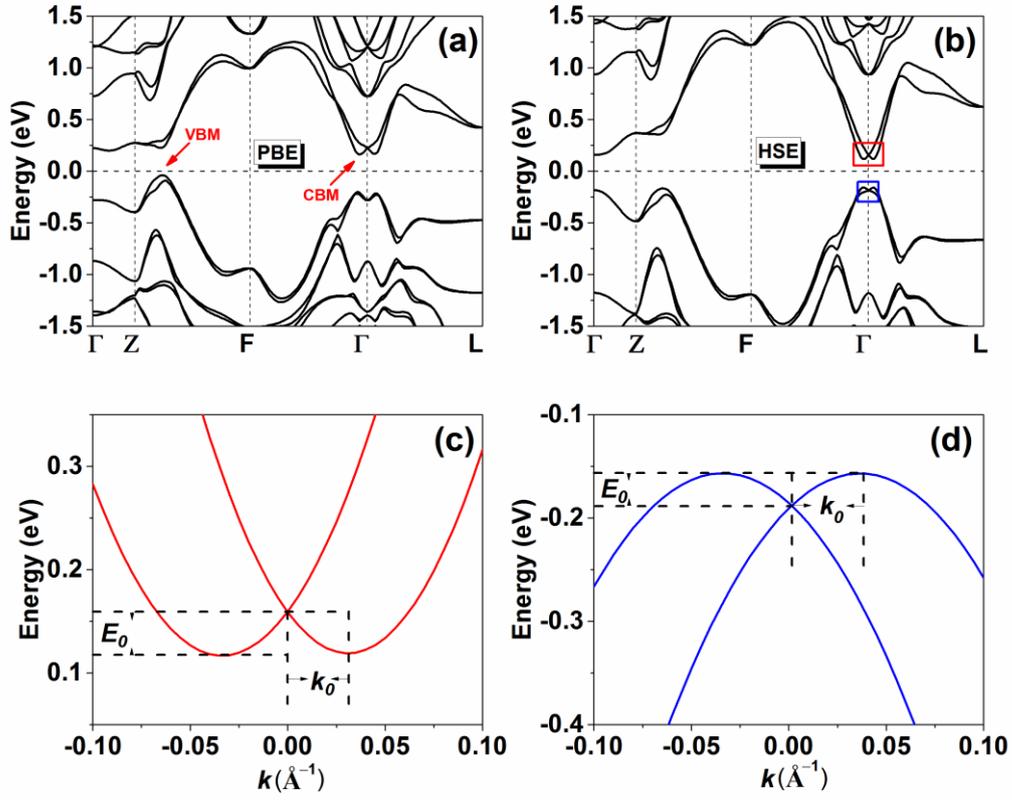

**Figure 2.** Band structures of BiSbSeTe$_2$ calculated with (a) PBE, and (b) HSE functionals. The enlarged views of the spin-splitting bands in (b) are shown in (c) and (d) for the conduction and valence bands, respectively.

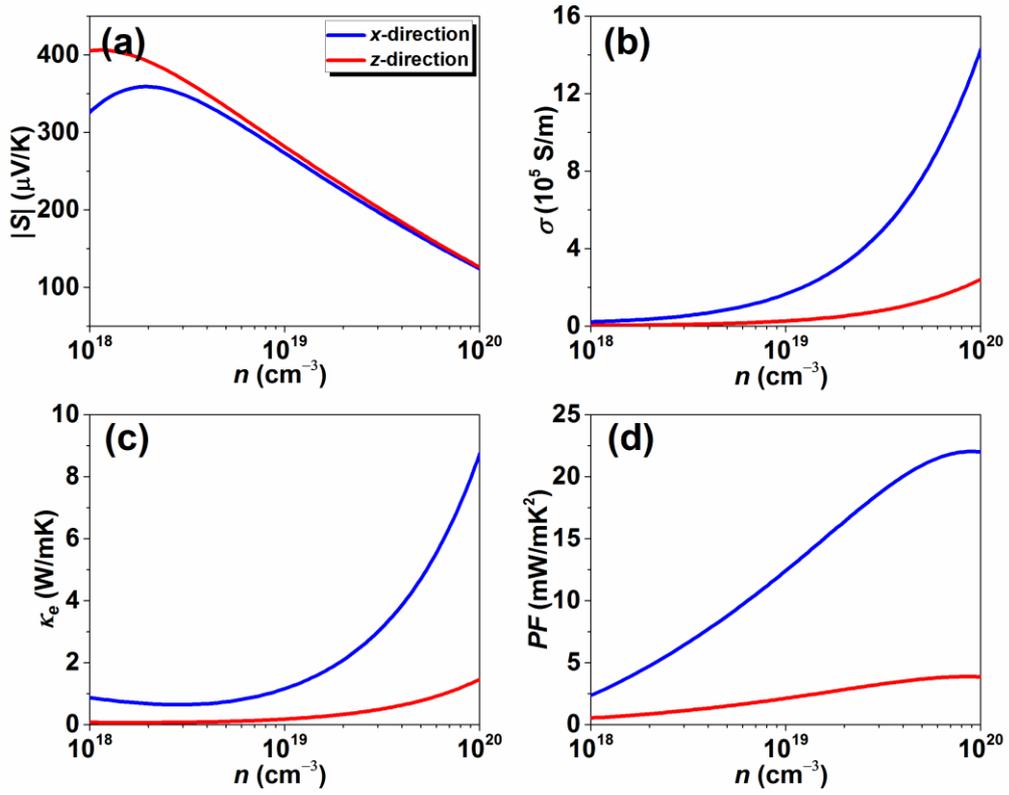

**Figure 3.** Room temperature electronic transport coefficients of BiSbSeTe$_2$: (a) the absolute values of the Seebeck coefficient, (b) the electrical conductivity, (c) the electronic thermal conductivity, and (d) the power factor, plotted as a function of hole concentration.

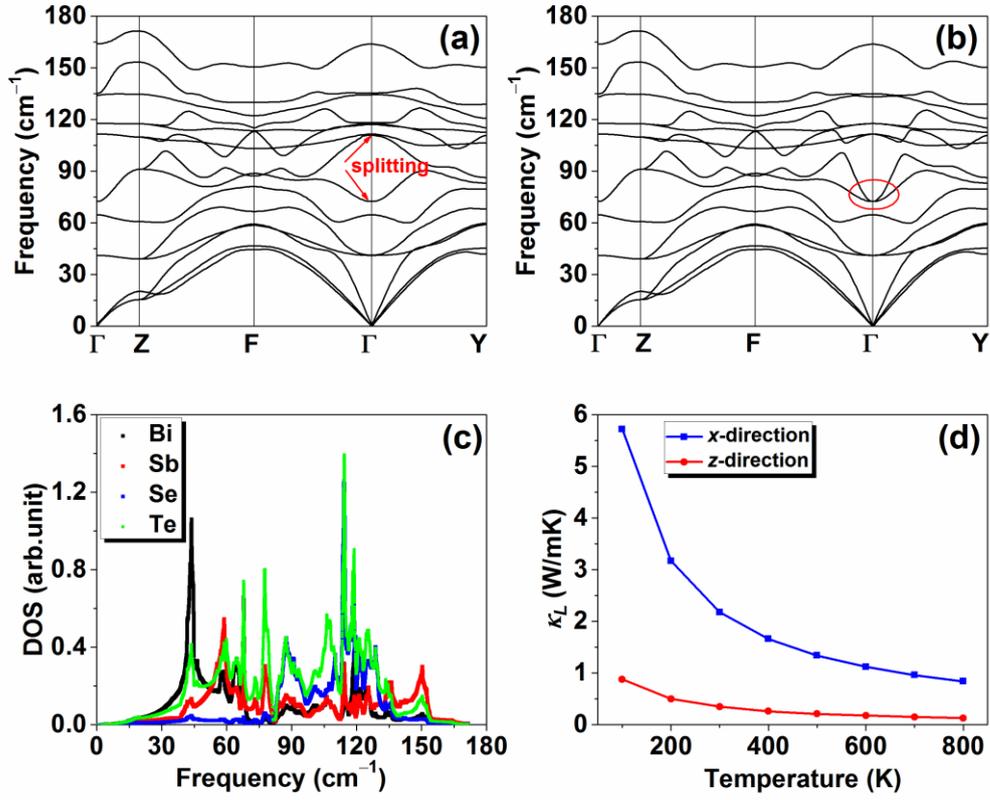

**Figure 4.** The phonon dispersion relations of BiSbSeTe$_2$ (a) with and (b) without the polarization field considered. (c) PDOS of BiSbSeTe$_2$. (d) The lattice thermal conductivity of BiSbSeTe$_2$ as a function of temperature.

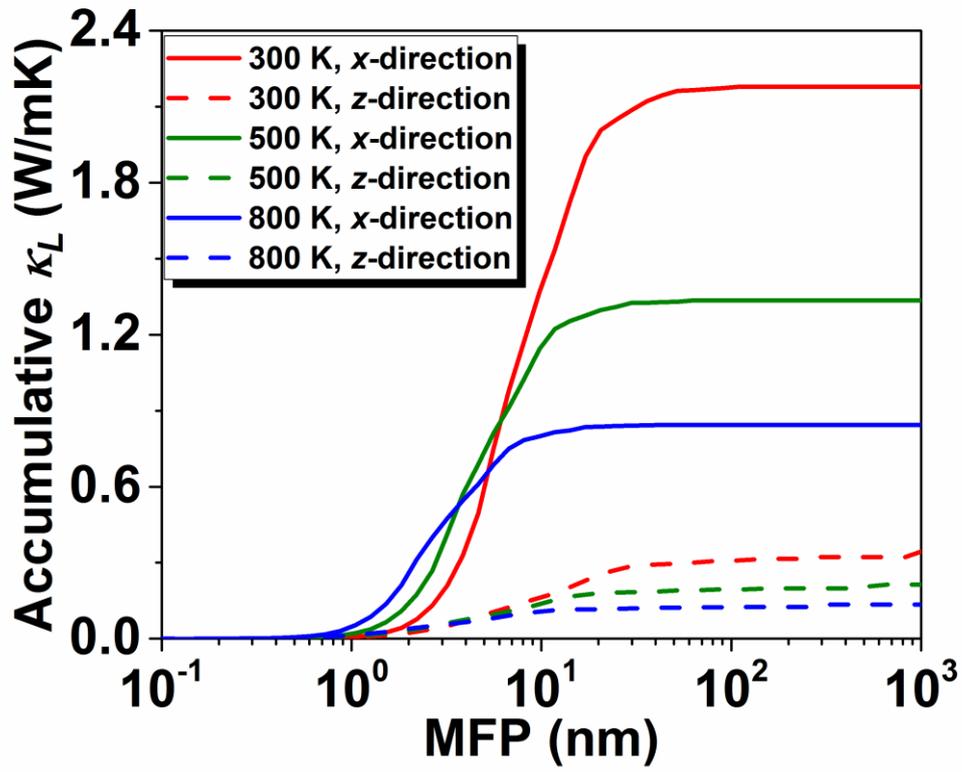

**Figure 5.** The accumulative lattice thermal conductivity of BiSbSeTe$_2$ with respect to phonon MFP at three typical temperatures.

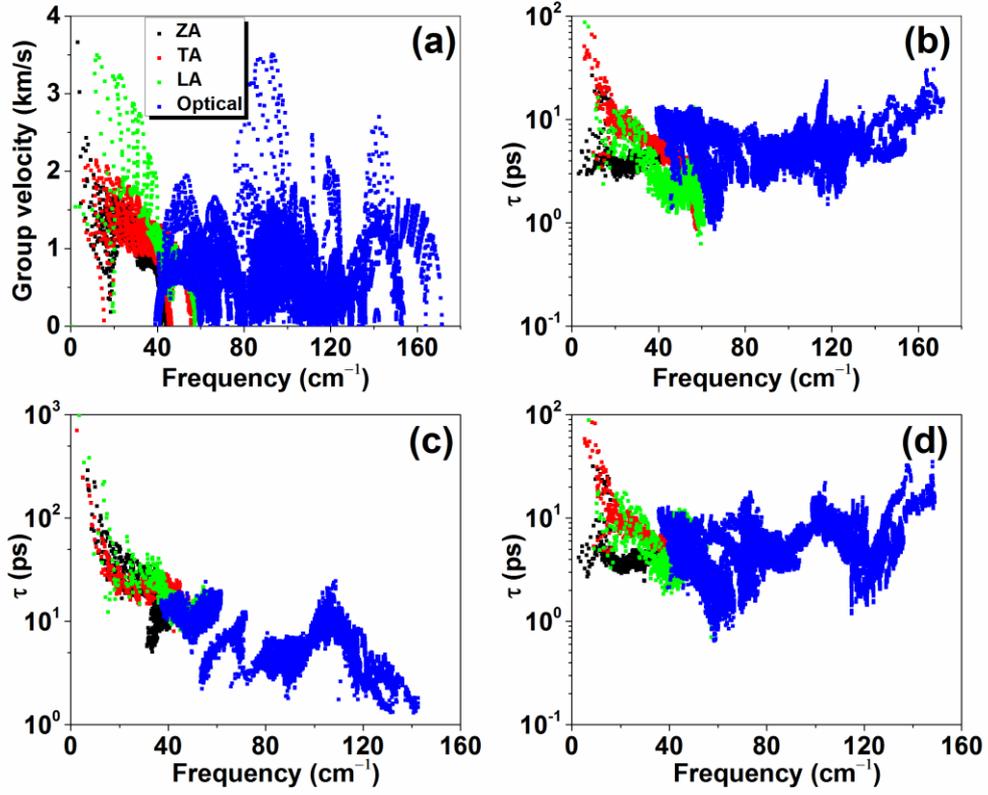

**Figure 6.** Room temperature (a) group velocity and (b) relaxation time of BiSbSeTe$_2$ as a function of frequency. For comparison, the relaxation time of Bi$_2$Te$_3$, and that for BiSbSeTe$_2$ with mass difference removed are shown in (c) and (d), respectively.

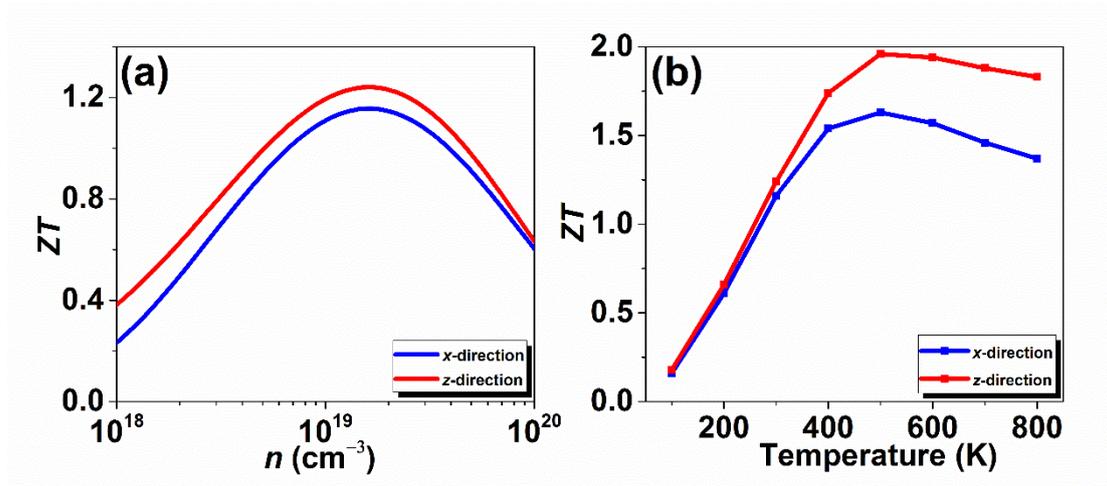

**Figure 7.** (a) Room temperature *p*-type *ZT* values of BiSbSeTe$_2$ as a function of carrier concentration along the *x*- and *z*-directions. (b) The temperature dependence of *ZT* value.